\documentclass[mnsc, nonblindrev]{informs3_preprint}

\OneAndAHalfSpacedXI

\usepackage{natbib}
 \bibpunct[, ]{(}{)}{,}{a}{}{,}%
 \def\bibfont{\small}%
\usepackage{hyperref}
\hypersetup{
	colorlinks,
	linkcolor=red,
  citecolor=blue,
  urlcolor=black
}
\usepackage{tabularx}
\usepackage{booktabs}
\usepackage{multicol}
\usepackage{multirow}
\usepackage{booktabs}
\usepackage{subfigure}
\usepackage{lmodern}
\usepackage{comment}

\newcolumntype{L}[1]{>{\raggedright\let\newline\\\arraybackslash\hspace{0pt}}m{#1}}
\newcolumntype{C}[1]{>{\centering\let\newline\\\arraybackslash\hspace{0pt}}m{#1}}
\newcolumntype{R}[1]{>{\raggedleft\let\newline\\\arraybackslash\hspace{0pt}}m{#1}}
\TheoremsNumberedThrough     
\ECRepeatTheorems

\EquationsNumberedThrough    

\begin{document}

\RUNAUTHOR{Che et al.}
\RUNTITLE{TimeMark: A Trustworthy Time Watermarking Framework}
\TITLE{TimeMark: A Trustworthy Time Watermarking Framework for Exact Generation-Time Recovery from AIGC}

\ARTICLEAUTHORS{%
\AUTHOR{Shangkun Che}
\AFF{Economics and Management School, Wuhan University, Wuhan, China\\
\EMAIL{csk105@whu.edu.cn}} 
\AUTHOR{Silin Du}
\AFF{School of Economics and Management, Tsinghua University, Beijing, China \\
\EMAIL{dsl21@mails.tsinghua.edu.cn}}
\AUTHOR{Ge Gao}
\AFF{School of Business Administration, Southwestern University of Finance and Economics, China \\ \EMAIL{gaoge@swufe.edu.cn}}
{\centering
Last Update: \today
}
} 

\ABSTRACT{
The widespread use of Large Language Models (LLMs) in text generation has raised growing concerns about intellectual property disputes. Watermarking techniques, which aim to embed meta information into AIGC, have the potential to serve as judicial evidence. However, existing LLM watermarking methods rely on statistical signals embedded in token distributions, leading to inherently probabilistic detection. Their reliability further degrades when encoding multi-bit information (e.g., timestamps), often falling far below 100\% accuracy. Moreover, these methods introduce detectable statistical patterns, making them vulnerable to forgery attacks, and also allow model providers to fabricate arbitrary watermarks.
To address these limitations, we introduce the concept of trustworthy watermark, which enables reliable watermark recovery with 100\% identification accuracy while resisting both user-side statistical attacks and provider-side forgery.
In this work, we focus on trustworthy time watermark which can serve as judicial evidence in intellectual property disputes. We integrate several cryptographic techniques and develop a novel trustworthy time watermarking framework.
Unlike existing approaches that directly encode time as the watermark payload, we embed time information into time-dependent secret keys under regulatory supervision, preventing provider-side forgery of arbitrary timestamps. 
The watermark payload is decoupled from time and instantiated as a randomly generated, non-stored bit sequence for each generation, eliminating statistical patterns and rendering statistical attacks ineffective. To address the unverifiability introduced by random payloads, we design a two-stage encoding mechanism that makes the recovered payload verifiable. Combined with error-correcting coding, this enables reliable recovery of generation time with theoretically perfect (100\%) identification accuracy.
Experimental results and theoretical analysis both demonstrate that the proposed framework meets the reliability requirements of judicial evidence. The framework provides a practical technical solution for addressing intellectual property disputes that are expected to arise frequently in future AIGC applications.}

\maketitle

\section{Introduction}
An increasing number of users now rely on Large Language Models (LLMs) to generate creative proposals, research reports and other forms of high-value AI-generated content (AIGC)\citep{bick2026rapid} \footnote{Consulting firm reports have also highlighted this trend, e.g., \url{https://www.mckinsey.com/capabilities/growth-marketing-and-sales/our-insights/how-generative-ai-can-boost-consumer-marketing} and \url{https://www.mckinsey.com/capabilities/quantumblack/our-insights/the-state-of-ai}} . 
As generative AI becomes embedded in knowledge work and commercial production, intellectual property disputes over authorship and ownership of AIGC are likely to become increasingly common. 
For example, different users may independently generate highly similar outputs with LLMs. Determining copyright ownership and identifying potential plagiarism in AIGC scenarios is particularly challenging. Unlike traditional forms of intellectual property evidence, which often provide reliable digital timestamps or version histories, LLM-generated content typically lacks trustworthy records of its creation time. Once users publish LLM-generated outputs online, only the publication time can be observed, while the actual generation time of the underlying idea remains unknown. This limitation can make certain judicial determinations extremely difficult, because the generation time of a document often constitutes a key evidentiary element in intellectual property disputes, particularly when courts must determine priority of creation or assess allegations of plagiarism \citep{wipo2021timestamp,siddharth2025time}.
In the copyright law, the key temporal point is not merely when a work is published, but when it is created and fixed. For example, under U.S. law, copyright subsists in original works of authorship ``fixed in any tangible medium of expression'' \citep{usc17_101,usc17_102}. UK law reflects a similar principle: copyright in literary, dramatic, and musical works arises only once the work is ``recorded, in writing or otherwise'' \citep{uk_cdpa_s3}. This legal structure shows that, in intellectual property disputes, the actual time of creation and fixation may be more important than the mere date of public dissemination \footnote{Timestamp is considered critical evidence in intellectual property disputes, which is also recognized by courts and experienced intellectual property lawyers, e.g., \url{https://english.bjinternetcourt.gov.cn/2022-04/02/c_531.htm} and \url{https://trustlynx.com/blog/post/Unquestionable-Authenticity-with-Timestamps\%20\%20}}.
Therefore, when used as judicial evidence for intellectual property disputes, it is not sufficient to merely determine whether a document was generated by an LLM. More importantly, one must be able to recover when the content was generated, since generation time is often indispensable for establishing priority, provenance, and ownership.

Under existing evidentiary frameworks for electronic evidence \citep{fre2024,coe_e_evidence,eidas_regulation}, a timestamp intended to support judicial fact-finding should satisfy several core requirements: (1) it should be sufficiently authentic, in the sense that the proponent can show that the record is what it claims to be; (2) it should be supported by a sufficiently reliable process or system capable of producing accurate results; (3) it should preserve the integrity of the associated data, so that later alterations are either prevented or rendered detectable; (4) it should be verifiable by the court or other parties; (5) it should arise from a regularly conducted business activity rather than being created ad hoc for litigation.
Translating these legal principles into practical system requirements, a judicially useful AIGC timestamping mechanism should therefore provide: (1) high evidentiary trustworthiness, meaning that the claimed generation time is accurate; (2) process reliability, meaning that each AIGC instance can be reliably associated with its timestamp and the service provider itself cannot forge the timestamp; (3) tamper resistance, meaning that tampered or forged texts cannot produce a valid or consistent timestamp under the verification procedure; (4) verifiability, such that any individual or institution can verify the timestamp based on the evidence, rather than relying on assertions by a particular service provider; (5) routine operational generation, so that the evidence is produced as part of the platform’s ordinary business workflow rather than created ad hoc for litigation.

LLM watermarking methods can embed target information into AIGC by introducing controlled bias into token sampling during the generation process. These methods operate natively within the text generation pipeline of the model. Because their algorithmic procedures are transparent and publicly specified, the embedded watermark can be verified by any institution with access to the algorithm and the corresponding token sampling mechanism. As a result, such methods naturally align with the requirements of ``verifiability'' and ``routine operational generation'' that are expected of a judicially useful AIGC timestamping mechanism.
Existing LLM watermarking methods have been developed primarily for generation detection, which is to determine whether a text was produced by an LLM \citep{kirchenbauer2023watermark,dathathri2024scalable}. Later studies moved beyond zero-bit detection and proposed multi-bit watermarking capable of embedding additional metadata, such as user identifiers or timestamps, into generated text \citep{yoo2024advancing,jiang2025stealthink}.
However, prior work on multi-bit watermarking shows that recovery accuracy decreases as message length increases, which is usually far below 100\% \citep{yoo2024advancing,jiang2025stealthink}. Beyond this problem, existing LLM watermarking methods suffer from more fundamental security and trustworthiness limitations. A key issue is that watermarks inevitably introduce statistical regularities into generated text. Although existing studies have proposed several strategies to reduce statistical signals \citep{jiang2025stealthink,diaa2024optimizing,zhang2024remark}, they typically reuse an identical watermark payload for the same user or within a fixed time window. This design exposes the watermarking mechanism to a stronger statistical attack: given unrestricted query access to a target model, an adversary can recover the biased token distribution corresponding to the current payload and thereby forge a text with the watermark \citep{jovanovic2024watermark}.
Moreover, existing approaches are limited to directly encoding time information as the watermark payload \citep{kirchenbauer2023watermark}. This design introduces an additional issue: the model provider itself can generate content with arbitrarily backdated timestamps, since no external mechanism constrains its ability to reconstruct past watermark payloads. As a result, watermark authenticity cannot be guaranteed.

Taken together, these limitations imply that current watermarking methods cannot simultaneously achieve reliable detection and strong anti-forgery guarantees. Consequently, they fail to meet the evidentiary standards required for judicial scenarios, where both correctness and resistance to both user-side and provider-side forgery are essential. 
These limitations directly violate the core requirements of a judicially useful AIGC timestamping mechanism introduced above, namely high evidentiary trustworthiness, process reliability, and tamper resistance. As a result, existing watermarking methods cannot satisfy the criteria of a trustworthy watermark and are therefore inadequate for serving as reliable judicial evidence.

For this reason, what is urgently needed is a novel trustworthy watermarking framework. In this work, we define a trustworthy time watermark as one that satisfies three key requirements. First, it should enable recovery of the embedded time information from generated text with 100\% accuracy. Second, it should prevent the model provider from forging watermarks with arbitrarily specified timestamps. Third, it should be robust against statistical attacks, thereby preventing users from forging valid time watermark through large-scale observation. 

To overcome the above limitations, this paper proposes a trustworthy time watermarking framework named TimeMark by integrating several ideas from cryptography and secure systems. 
First, unlike existing approaches, we fundamentally depart from directly encoding information into the watermark. Instead, we encode the generation time into time-dependent keys, which are derived through a one-way key evolution function and it is unable to derive past keys from the current one. The secret keys are securely managed under regulatory supervision via a Hardware Security Module (HSM). This design prevents service providers from forging watermarks corresponding to arbitrary timestamps, as past keys cannot be reconstructed without authorized access.
Second, during the watermark encoding process, the secret key is not directly embedded as the watermark payload into the generated text. Rather, the watermark payload consists of a randomly generated bit sequence, which is newly created for each generation and never stored. This randomized payload construction eliminates any persistent statistical patterns in the generated text, thereby ensuring that, in principle, statistical attacks cannot be used to forge valid time watermarks. To support this design, we further introduce a novel two-stage encoding mechanism with different greenlist-generation logics.
Finally, by incorporating an error-correcting coding technique called Bose–Chaudhuri–Hocquenghem codes (BCH), our framework achieves reliable recovery of the payload and ensures perfect (100\%) identification accuracy of the generation time.

In conclusion, the TimeMark framework fulfills all essential requirements of a trustworthy time watermark, achieving accurate generation-time identification while preventing both user-side and provider-side forgery. These strong guarantees make it a promising candidate for reliable judicial evidence in intellectual property disputes.

This work makes contributions on both the problem formulation and methodological design. On the one hand, we identify an important yet largely overlooked problem in the literature on LLM watermarking, namely the need for trustworthy time watermark in the context of AIGC-related intellectual property disputes, which are likely to become increasingly prevalent. In such settings, watermarking methods must satisfy a set of stringent requirements to be considered judicially useful, yet these requirements have received limited attention in existing studies, thereby calling for fundamentally new designs.
On the other hand, from a methodological perspective, the key innovation of this work lies in decoupling the timestamp information from the watermark payload. Instead of directly encoding timestamp information as the payload, we employ a randomly generated sequence as the payload and bind the timestamp to a strongly regulated time-dependent secret key. This design effectively prevents provider-side forgery and user-side statistical attacks.

\section{Related Work}

\subsection{Intellectual Property Law and Evidentiary Frameworks}

The intellectual property disputes over AIGC should be understood against the broader background of copyright law and electronic evidence.  
In the United States, copyright protection subsists in ``original works of authorship fixed in any tangible medium of expression'' \citep{usc17_102,usc17_101}. The U.S. Copyright Office has further taken the position that existing copyright doctrine remains applicable to generative-AI outputs, while continuing to study unresolved questions concerning AI-assisted authorship \citep{usco_ai_overview,usco_ai_part2,usco_ai_part3}. In the United Kingdom, copyright law likewise remains grounded in the Copyright, Designs and Patents Act 1988, while the government continues to evaluate possible reforms concerning copyright and AIGC through consultation and evidence-gathering \citep{uk_cdpa_s3,uk_copyright_ai_consultation,uk_copyright_ai_progress}. 
The European Union exhibits a closely related regulatory framework. On the copyright side, the Digital Single Market Directive remains a central instrument because its framework is directly relevant to disputes over generative models \citep{eu_dsm_directive}. At the same time, the EU Artificial Intelligence Act now imposes transparency- and copyright-related obligations on providers of general-purpose AI models, including obligations relevant to compliance policies and public summaries of training data \citep{eu_ai_act,eu_gpai_code,eu_gpai_qa}. China similarly combines its general copyright law with newer AI-governance instruments. Copyright questions remain rooted in the Copyright Law of the People's Republic of China, while the Interim Measures for the Management of Generative Artificial Intelligence Services and the 2025 Measures for the Labeling of AI-Generated Synthetic Content introduce additional governance obligations concerning generative AI services and the labeling of AIGC \citep{china_copyright_law,china_genai_measures,china_ai_label_measures}. 

These frameworks collectively suggest that future disputes over AIGC are likely to involve not only questions of authorship and ownership, but also disputes over priority, derivation, access, and copying. In such cases, the ability to establish when a disputed text was actually generated may become highly probative. This is particularly true because copyright law in major common-law jurisdictions does not treat public release as the sole decisive event. Rather, the legally significant temporal point is often the time of creation and fixation. Under U.S. law, copyright subsists upon fixation rather than publication \citep{usc17_101,usc17_102}; under UK law, copyright does not subsist unless and until the work is ``recorded, in writing or otherwise'' \citep{uk_cdpa_s3}. 

From the perspective of evidentiary doctrine, these copyright concerns must be translated into requirements for judicially useful time evidence. 
U.S. evidence law requires authentication: the proponent must produce evidence sufficient to support a finding that the item is what the proponent claims it is \citep{fre901}. This supports a first requirement for AIGC time evidence: \emph{it should be sufficiently authentic, in the sense that the proponent can show that the record is what it claims to be}. 
U.S. law also specifically recognizes evidence describing a process or system and showing that it produces an accurate result \citep{fre901,fre902}. This supports a second requirement: \emph{it should be supported by a sufficiently reliable process or system capable of producing accurate results}. 
A third requirement follows from the law's concern with data integrity. The Council of Europe has emphasized that electronic evidence should be evaluated with particular regard to admissibility, authenticity, accuracy, and integrity \citep{coe_e_evidence}. In EU law the requirements for qualified electronic timestamps demand a link to data in a manner reasonably precluding undetectable change \citep{eidas_regulation}. This supports the proposition that judicially useful AIGC time evidence \emph{should preserve the integrity of the associated data, so that later alterations are either prevented or rendered detectable}. 
A fourth requirement is verifiability. Rule 902's certification-and-notice structure is designed to make electronic records open to adversarial testing, and the Council of Europe guidelines similarly contemplate judicial and expert scrutiny of electronic evidence where necessary \citep{fre902,coe_e_evidence}. For AIGC disputes, this implies that a timestamping or provenance mechanism \emph{should not rely on inaccessible, purely internal assertions, but should instead permit meaningful checking by courts, experts, and adverse parties}. 
Finally, a fifth requirement is that such evidence \emph{should arise from a regularly conducted business activity rather than being created ad hoc for litigation}. Rule 803(6) admits records of a regularly conducted activity when they are made at or near the time by someone with knowledge, kept in the course of regularly conducted activity, and made as a regular practice \citep{fre803}. In the AIGC context, this means that the time evidence should generally be generated natively as part of the model provider's ordinary operational workflow. 

Taken together, these legal frameworks motivate a concrete evidentiary design target for future AIGC systems. These requirements do not by themselves resolve all questions of copyright ownership or infringement, but they provide a principled evidentiary framework for making generation-time evidence more judicially useful in future disputes involving AIGC.
These legal principles can be systematically translated into practical system requirements as shown in Table~\ref{tab:legal_to_system_requirements}.

\begin{table}[t]
\centering
\caption{From legal evidentiary principles to practical requirements for AIGC timestamping mechanisms.}
\label{tab:legal_to_system_requirements}
\begin{tabularx}{\linewidth}{p{0.24\linewidth} p{0.24\linewidth} X}
\toprule
\textbf{Legal requirement} & \textbf{Practical requirement} & \textbf{Explanation} \\
\midrule

Sufficient authenticity &
High evidentiary trustworthiness &
The claimed generation time should accurately correspond to the specific record offered as evidence, so that the record is what the proponent claims it to be. \\

Reliable process or system capable of producing accurate results &
Process reliability &
Each AIGC output should be reliably bound to its timestamp through a trustworthy generation procedure, and the service provider itself should not be able to forge this association arbitrarily. \\

Preservation of data integrity &
Tamper resistance &
Tampered, manipulated, or forged texts should not produce a valid timestamp under the verification procedure. \\

Verifiability by the court or other parties &
Verifiability &
Timestamp validation should rely on inspectable evidence and a public verification procedure, rather than on unilateral assertions by a particular service provider. \\

Arising from a regularly conducted business activity &
Routine operational generation &
The evidence should be generated natively as part of the platform's ordinary operational workflow, rather than being created retrospectively for litigation. \\

\bottomrule
\end{tabularx}
\end{table}

\subsection{Timestamping Methods}

Existing timestamping methods for digital content can be roughly divided into four categories: cryptographic trusted timestamping, append-only logging and provenance systems, blockchain-based timestamp anchoring, and LLM watermarking. 

For classical cryptographic timestamping, \cite{haber1991timestamp} proposed mechanisms for proving that a digital document existed at or before a given time without revealing its contents. This line of work was later standardized in the Time-Stamp Protocol (TSP), where a trusted time-stamping authority issues a signed token over the hash of a document \citep{rfc3161}. Closely related approaches appear in secure logging and provenance systems, which protect the temporal order and integrity of records through append-only or forward-secure constructions \citep{ma2009securelogging}. More recently, content provenance standards such as C2PA provide signed metadata structures for recording the origin and modification history of digital assets \citep{c2pa2025}. 
Another widely discussed direction is blockchain-based timestamping, which anchors document hashes or metadata to public ledgers in order to obtain decentralized immutability guarantees \citep{gipp2017blockchain,meng2022blockchain}.

However, these methods are not suitable for timestamping AIGC. 
The fundamental limitation of these approaches lies in the nature of how LLM-generated text is produced and delivered. Unlike traditional digital documents, which are created, stored, and transmitted as structured files with well-defined storage media, AIGC is typically generated on demand and transmitted directly to users as plain text over network interfaces. As a result, there is no persistent document container at the time of generation to which metadata or timestamps can be intrinsically attached.
In principle, a user could subsequently encapsulate the generated text into a digital file and apply the above timestamping techniques. However, such a timestamp would only certify the time at which the file was created or registered, rather than the actual time at which the text was generated by the model. This temporal mismatch fundamentally undermines the evidentiary value of the timestamp in judicial contexts. It violates a key requirement for trustworthy electronic evidence that the evidence should be generated natively as part of a regularly operational process. 
In addition, append-only logging and provenance systems rely on platform-maintained logs or metadata, and for plain text it remains vulnerable to format conversion or metadata stripping. Thus they do not provide a robust, text-intrinsic proof of generation time. 
Blockchain research continues to identify low throughput, high latency, and storage overhead as major scalability bottlenecks \citep{bulgakov2024scalability,chen2024comprehensive}. Moreover, placing large volumes of AIGC with metadata into external infrastructures introduces additional privacy concerns. Therefore, the above solutions are unlikely to be practical as a universal foundation for trustworthy AIGC timestamp.

Finally, in the context of language models, watermarking methods embed detectable signals directly into generated text by biasing token sampling during generation \cite{kirchenbauer2023watermark,liu2023upv,ren2024semamark}. LLM watermarking is closer to the target setting because it works inside the decoding process and embeds evidence directly into the generated text \cite{kirchenbauer2023watermark}. Although existing LLM watermarking methods satisfy the requirements of ``verifiability'' and ``routine operational generation'' in Table~\ref{tab:legal_to_system_requirements}, they fail to meet other key evidentiary requirements, as will be discussed in the following subsection.

\subsection{LLM Watermarking}

LLM watermarking has primarily followed a decoding-time statistical watermarking paradigm, where a secret-key-dependent bias is injected into token sampling during generation and later detected through statistical testing. A representative example is the green-list framework, which biases the logits of a pseudo-random subset of tokens and then uses hypothesis testing to determine whether the resulting text carries a watermark \citep{kirchenbauer2023watermark}. 
SynthID-Text advances this line of work toward practical deployment, offering efficient detection, negligible latency, and no training modifications \citep{dathathri2024scalable}. In these works, watermark payloads are embedded during text generation and inferred through aggregate statistical evidence.
Beyond decoding-time methods, alternative paradigms such as semantic-space watermarking have been proposed. However, this approach requires additional training and incur higher deployment complexity and more complex detection procedures \citep{zhang2024remark}.
Most studies focuse on zero-bit watermarking, whose goal is simply to determine whether a text was generated by an LLM. Recent work has begun to explore multi-bit watermarking, where the generated text carries additional meta information. \citet{wang2024towards} extend conventional single-bit watermarking to a multi-bit codable framework, where auxiliary information such as user identities can be embedded and later recovered through balanced vocabulary partitioning. \citet{yoo2024advancing} introduce a position allocation mechanism that distributes different portions of a payload across token positions, thereby enabling more reliable multi-bit message recovery. The shift from zero-bit to multi-bit watermarking marks an important transition from mere AI-text detection toward fine-grained provenance encoding and traceability.

A major line of research seeks to improve robustness against modification, often discussed under the scrubbing attacks, where an adversary rewrites, edits, or paraphrases the watermarked text in order to erase the signal. For example, DiPmark improves robustness by combining distribution-preserving reweighting with statistically grounded detection, showing resilience under moderate editing and paraphrasing \citep{wu2024resilient}. SemaMark strengthens resistance to paraphrasing by replacing token-level hashing with semantically informed partitioning, thereby making the watermark seed more stable under meaning-preserving rewrites \citep{ren2024robust}. The literature also makes clear that robustness remains fundamentally limited: strong paraphrasing, translation, or sufficiently adaptive rewriting can still sharply reduce detectability \citep{rastogi2024revisiting}.

Another direction concerns unforgeability (i.e., resistance to spoofing attacks), namely whether an adversary can fabricate a watermark in non-watermarked text. For example, publicly detectable watermarking addresses this problem by incorporating cryptographic signatures and public verification, so that watermark validity does not depend on trusting the detector alone \citep{fairoze2025publicly}. However, they lack a mechanism to securely bind generation time, leaving them vulnerable to provider-side forgery and time-shift attacks. Moreover, their direct encoding of semantic payloads (e.g., timestamps) introduces structural patterns and increases vulnerability under attacks. \citet{christ2024undetectable} formalize a related notion through undetectable watermarking, showing how watermarking can be made computationally indistinguishable without the secret key, while also clarifying the security limits of such constructions. The literature also highlights intrinsic trade-offs: stronger robustness may enlarge the attack surface for spoofing or reverse engineering, and black-box attacks show that some statistical watermark designs can be inferred or potentially imitated under realistic adversarial access \citep{rastogi2024revisiting}.

A further group of studies focuses on the trade-off between text quality and detectability. Unbiased watermarking aims to preserve the original output distribution while still enabling detection, thereby minimizing perceptible degradation in text quality \citep{hu2024unbiased}. Similarly, distribution-preserving methods attempt to maintain fluency and semantic fidelity while keeping the watermark statistically recoverable \citep{wu2024resilient}. These methods are especially important because practical deployment requires not only detectability, but also minimal distortion to model behavior and user experience. 

In summary, existing improvements can be broadly grouped into three directions: enhancing robustness against scrubbing attacks, strengthening unforgeability against spoofing attacks, and improving the trade-off between detectability and text quality.
Existing watermarking methods are not well suited for use as judicial evidence in intellectual property disputes, especially for generation-time verification. Prior research has rarely examined what requirements a watermark must satisfy to serve as reliable judicial evidence, and has focused more on resistance to scrubbing attacks than on spoofing attacks. Yet, in our setting, spoofing is the more critical threat: watermark removal only causes missing evidence, whereas watermark forgery produces false evidence and fundamentally undermines credibility. Moreover, existing methods often embed the target information itself, such as the timestamp, in watermark payloads, which means that a service provider with access to the embedding mechanism could generate text with any arbitrarily specified time. The limited diversity of watermark payloads also makes statistical imitation easier, and current methods generally do not guarantee perfectly accurate recovery of target information, making them insufficiently reliable for judicial evidence. Therefore, a new time watermarking framework is needed to achieve the level of trustworthiness required for judicial evidence.

\section{Preliminaries}
\subsection{LLM Watermarking Basics}
LLM watermarking aims to embed hidden signals into model-generated text so that the text can later be verified, identified, or decoded. Existing watermarking methods can be broadly classified in two ways. First, based on the amount of information carried by the watermark, they can be divided into {zero-bit} and {multi-bit} watermarking. {\bf Zero-bit watermarking} embeds only binary information to indicate the presence of a watermark, while {\bf multi-bit watermarking} encodes richer payloads such as user identity and metadata. Second, according to how the watermark is embedded, methods can be divided into generation-time approaches that intervene in the text generation process and alternative paradigms such as semantic-space or training-based watermarking. Among these directions, the most widely studied and practically influential line is {decoding-time token-level watermarking} \citep{kirchenbauer2023watermark,dathathri2024scalable}. Since our framework also embeds signals by modifying the token sampling process during generation, this subsection focuses on that paradigm.

In decoding-time token-level watermarking, the watermark is embedded by slightly perturbing the next-token distribution during autoregressive text generation. Rather than forcing the model to output predetermined tokens, the watermarking algorithm introduces a small bias into sampling, so that the generated text still remains fluent and semantically natural while carrying a hidden statistical trace \citep{kirchenbauer2023watermark}. This paradigm can be viewed through an encoder–decoder framework. The encoder injects the watermark by modifying the token sampling process, while the decoder analyzes the generated text to detect the presence of the watermark or recover the embedded information.

At each generation step, the model first produces a probability distribution over the vocabulary for the next token. The watermarking mechanism then partitions the vocabulary into two disjoint subsets. These two subsets are referred to as the \emph{greenlist} (preferred set) and the \emph{redlist} (non-preferred set). 
{This token partition is constructed dynamically using a pseudo-random procedure conditioned on the current prefix, and must be fully reproducible so that it can be reconstructed during decoding.} The model's sampling distribution is then modified so that tokens in the preferred set become slightly more likely to be selected. Formally, let $\mathcal{V}$ be the vocabulary. For position $i$, let $c_i$ denote the sequence of tokens preceding the current position $i$ (i.e., the context) and $p_i(v\mid c_i)$ denote the original next-token distribution of the LLM for token \(v \in \mathcal{V}\). The vocabulary is partitioned into two sets, i.e., the greenlist $\mathcal{G}_i$ and the redlist $\mathcal{R}_i$. 

In zero-bit watermarking, the encoder aims to embed a detectable watermark signal. The encoder follows a fixed rule across all positions: it consistently biases the sampling distribution toward the greenlist.
\[
\tilde{p}(v\mid c_i) \propto 
\begin{cases}
p(v\mid c_i)\exp(\delta), & v \in \mathcal{G}_i,\\
p(v\mid c_i), & v \in \mathcal{R}_i,
\end{cases}
\]
where \(\delta > 0\) controls the watermark strength. 
In contrast, multi-bit watermarking aims to encode a payload consisting of multiple bits. To achieve this, the encoder specifies a \emph{bit-allocation rule}, which determines which message bit is assigned to each token position. For example, under a fixed-length allocation rule, every consecutive 10 token positions may be assigned to encode one payload bit, so that positions (1) to (10) encode the first bit, positions (11) to (20) encode the second bit, and so on. 
Given such a rule, the token partition can then be used in a position-dependent manner. The encoder selects which subset (greenlist or redlist) to favor at each position according to the payload being embedded. For example, it may favor the greenlist when the assigned bit is 1 and the redlist when the assigned bit is 0. In this way, different token positions jointly encode different parts of the payload, and the final text carries a recoverable bit sequence.

During the decoding process, the first step is to reconstruct the token partition for each position by applying the identical pseudo-random procedure.
In zero-bit settings, the decoder examines the generated tokens and counts how often they fall into the greenlist. Since the encoder consistently prefer the greenlist, a watermarked text is expected to contain a higher-than-random proportion of tokens in the greenlist. The decoder therefore performs a statistical test to assess whether the observed frequency significantly exceeds the expected baseline. Based on this test, it outputs a binary decision.

In multi-bit settings, with the bit-allocation rule, the decoder can determine which message bit each position corresponds to. 
It then infers the encoded bit at each position based on whether the generated token falls into the greenlist or the redlist. Specifically, if the generated token falls into the greenlist, it provides evidence that the corresponding bit is 1; otherwise, it provides evidence that the bit is 0. 
The decoder then aggregates the evidence from all positions assigned to the same bit and determine the final bit value typically through a majority vote. In this way, the decoder reconstructs the entire payload by combining information across multiple token positions.

\subsection{Limitations of Existing Watermarking for Trustworthy Timestamping}

A straightforward approach to generation-time verification is to directly apply multi-bit watermarking and encode the timestamp as the watermark payload. However, this naive design faces several fundamental limitations. First, existing watermarking methods do not guarantee perfectly accurate recovery of the embedded information. Multi-bit watermarking typically relies on aggregating weak statistical signals across token positions, which makes decoding inherently probabilistic. Even small decoding errors can lead to incorrect timestamp recovery, which is unacceptable in evidentiary settings. Therefore, we do not encode the time-related information in its raw form. We transform the it into a redundant codeword using an error-correcting code so that the meaningful payload can still be exactly recovered even when some watermark bits are decoded incorrectly.

Second, the service provider controls the encoding process, it can generate text with arbitrarily specified timestamps. In other words, there is no inherent mechanism preventing the provider from backdating or postdating content. To address this, the encoding process should rely on a time-evolving key mechanism in which historical keys are not held by the model provider but are managed by an external trusted authority, and are accessed only during timestamp verification.

Third, directly encoding timestamps as watermark payloads introduces a structural vulnerability that enables statistical imitation attacks. To make this concrete, we describe a representative attack under a realistic adversarial setting.

Consider a standard multi-bit token-level watermarking scheme. Let $t$ denote the timestamp of a generated document, and let $b(t)=(b_1(t),\dots,b_m(t))\in \{0,1\}^m$ be the bit sequence obtained by encoding this timestamp. A fixed bit-allocation rule $a(i)\in\{1,\dots,m\}$ assigns each token position $i$ to one payload bit, meaning that position $i$ is used to encode bit $b_{a(i)}(t)$. The watermarking mechanism uses the secret key to seed a pseudo-random procedure that partitions the vocabulary at each position.

We assume that the adversary has black-box access to the watermarked LLM and knows the timestamp-to-bit encoding rule $b(\cdot)$, the bit-allocation rule $a(\cdot)$, and the watermarking algorithm, but does not know the secret key. This is the standard security setting in which the mechanism is public and only the key is hidden~\citep{jovanovic2024watermark}. The adversary’s goal is to generate a forged text that causes the decoder to recover a target timestamp $t^\star$, without following the genuine encoding process. The attack proceeds in four steps. 
\begin{enumerate}
    \item {\bf Data Collection}. The adversary collects a set of watermarked documents $\{(x^{(n)},t^{(n)})\}_{n=1}^N$, where $x^{(n)}=(x_1^{(n)},x_2^{(n)},x_3^{(n)},\dots)$ is a generated document and $t^{(n)}$ is its timestamp label. Since the timestamp-to-bit mapping is known, the adversary can compute the payload of each collected document as $b(t^{(n)})=(b_1(t^{(n)}),\dots,b_m(t^{(n)}))$.
    
    \item {\bf Dataset Convertion}.
    For each document $x^{(n)}$ and each token position $i$, the adversary constructs the context $c_i^{(n)}=(x_1^{(n)},\dots,x_{i-1}^{(n)})$ and assigns the label $y_i^{(n)}=b_{a(i)}(t^{(n)})$, which is the bit value encoded at that position. This transformation is feasible since both $b(\cdot)$ and $a(\cdot)$ are known.
    Then each observed token can be rewritten as a triplet $(c_i^{(n)},x_i^{(n)},y_i^{(n)})$.
    
    \item {\bf Model Training}. Using these triplets, the adversary trains a surrogate classifier to estimate the statistical association between token choices and encoded bit values. Concretely, for each sample $(c_i^{(n)},x_i^{(n)},y_i^{(n)})$, the adversary treats the context--token pair $(c_i^{(n)},x_i^{(n)})$ as the input and the bit label $y_i^{(n)}\in \{0,1\}$ as the supervision signal. 
    Let $h(y\mid c,v)$ denote a parametric classifier that outputs the probability that candidate token $v$ in context $c$ is associated with bit value $y$. The adversary trains $h$ on the collected samples with a standard binary classification objective. After training, the adversary converts the classifier output into a surrogate score
    $$s(c,v)=\log \frac{h(1\mid c,v)}{h(0\mid c,v)}$$
    which captures the relative tendency of token $v$ under context $c$ with respect to the two encoded bit values.  A positive and larger value of $s(c,v)$ indicates that, under context $c$, choosing token $v$ is more indicative of encoding bit 1 rather than bit 0.
    \item {\bf Forged Document Generation}. After obtaining the surrogate score, the adversary utilizes it to generate a forged document for a target timestamp $t^\star$. Let $b(t^\star)=(b_1(t^\star),\dots,b_m(t^\star))$ be the target bit sequence. At each generation step $i$, suppose the current forged prefix at position $i$ is $\tilde c_i=(\tilde x_1,\dots,\tilde x_{i-1})$. The adversary first determines target bit assigned to this position $y_i^\star=b_{a(i)}(t^\star)$. It then obtains a base next-token distribution $p(v\mid \tilde c_i)$ from a language model and modifies this distribution using the learned surrogate score $s(\tilde c_i,v)$. A natural implementation is to define a reweighted decoding distribution
    $$\tilde q_i(v\mid \tilde c_i,y_i^\star)\propto p(v\mid \tilde c_i)\exp\big(\lambda(2y_i^\star-1)s(\tilde c_i,v)\big) = \begin{cases}
p(v\mid \tilde{c}_i)\exp(\lambda s(\tilde c_i,v)), & y_i^\star = 1\\
p(v\mid \tilde{c}_i)\exp(-\lambda s(\tilde c_i,v)), & y_i^\star = 0
\end{cases}$$
    where $\lambda>0$ controls the attack strength. When $y_i^\star=1$, the factor $\exp(\lambda s(\tilde c_i,v))$ increases the probabilities of tokens with positive surrogate scores and decreases the probabilities of tokens with negative surrogate scores. In contrast, when $y_i^\star=0$, the factor $\exp(-\lambda s(\tilde c_i,v))$ reverses this preference. The adversary then samples the next token from $\tilde q_i(\cdot\mid \tilde c_i,y_i^\star)$ and continues to the next position. As a result, the forged text exhibits token choices similar to those that would be produced by genuine multi-bit watermark encoding. Since the decoder recovers each bit by aggregating token-level evidence across the positions assigned to that bit, such a forged text can induce the decoder to output the target timestamp $t^\star$ with non-negligible probability. 
\end{enumerate} 

Under a similar setting, ~\citet{jovanovic2024watermark} show that an adversary can learn an approximate surrogate watermark function with about 30,000 API queries and then generate forged texts that fool the watermark detector with an average success rate of over 80\%. To avoid any statistical attacks, we propose a framework to encode a fully random payload, while timestamp verification is realized through a separate mechanism-level design. Such a random payload can prevent adversaries from learning stable bit-dependent statistical patterns from large collections of watermarked texts.

\section{Problem Formulation}

Let $\mathcal{D}_t$ denote the set of all documents generated by a large language model (LLM) at time $t$. We consider a watermarking scheme that embeds temporal information into generated text. Specifically, let $\mathcal{M}_{\text{Enc}}$ and $\mathcal{M}_{\text{Dec}}$ denote the watermark encoding and decoding functions, respectively.

Given a document $d \in \mathcal{D}_t$, the encoder produces a watermarked document:
\begin{equation}
d^t = \mathcal{M}_{\text{Enc}}(t, d),
\end{equation}
where $t$ is the target timestamp to be embedded.

The decoder takes a document as input and outputs the recovered timestamp:
\begin{equation}
\hat{t} = \mathcal{M}_{\text{Dec}}(d^t).
\end{equation}

\begin{definition}
A watermarking scheme $(\mathcal{M}_{\text{Enc}}, \mathcal{M}_{\text{Dec}})$ is said to be \emph{trustworthy} if it satisfies the following properties:
\begin{itemize}
    \item \textbf{Correctness:} For any $t$ and any $d \in \mathcal{D}_t$,
    \begin{equation}
    \mathcal{M}_{\text{Dec}}(\mathcal{M}_{\text{Enc}}(t, d)) = t.
    \end{equation}

    \item \textbf{Unforgeability (Provider-side):} At time $t$, the LLM service provider cannot generate a document that encodes an arbitrary timestamp $t' \neq t$, i.e., it is infeasible to produce $d^{t'}$ such that
    \begin{equation}
    \mathcal{M}_{\text{Dec}}(d^{t'}) = t'.
    \end{equation}

    \item \textbf{Unforgeability (User-side):} Even with unrestricted access to the LLM at time $t$, a user cannot construct a document $d_{\text{user}}$ such that
    \begin{equation}
    \mathcal{M}_{\text{Dec}}(d_{\text{user}}) = t,
    \end{equation}
    unless it is legitimately generated by the watermarking mechanism.
\end{itemize}
\end{definition}

The goal of this work is to design a \emph{trustworthy time watermarking model} $\mathcal{M}$ with encoder $\mathcal{M}_{\text{Enc}}$ and decoder $\mathcal{M}_{\text{Dec}}$ for LLM-generated text and $\mathcal{M}$ satisfies the three properties in Definition 1. 

\section{Method}

\begin{figure}[!htbp]
    \centering
    \includegraphics[width=1\linewidth]{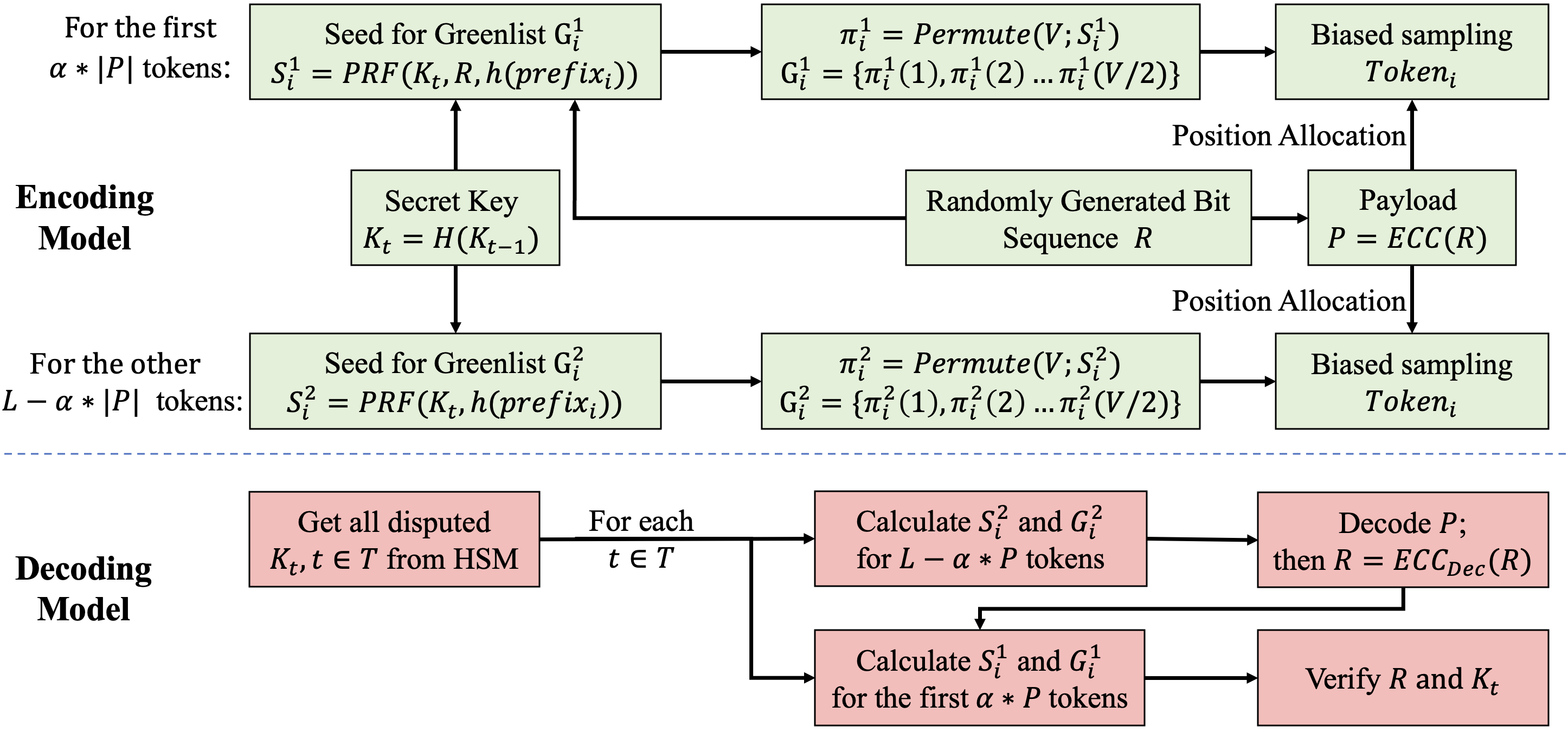}
    \caption{The Proposed Trustworthy Time Watermarking Framework}
    \label{fig1}
\end{figure}

Figure~\ref{fig1} illustrates the overall pipeline of the proposed trustworthy time watermarking framework, which consists of an \emph{encoding model} for watermark injection and a \emph{decoding model} for generation-time recovery and verification.

\subsection{Encoding Model}

We divide time into discrete windows indexed by $t$, where the granularity of each window is configurable (e.g., one minute). For each time window $t$, the watermarking system uses a secret key $K_t$ derived from a one-way hash chain:
\begin{equation}
K_t = H(K_{t-1}),
\end{equation}
where $H(\cdot)$ is a cryptographic hash function such as SHA-256. Due to the one-way property of $H(\cdot)$, past keys cannot be reconstructed from the current key. 
The key is securely stored in a Hardware Security Module (HSM), a specialized hardware device designed for secure key generation, storage, and cryptographic operations with strict access control. HSMs are widely deployed in highly security-sensitive environments such as banking systems, government institutions, and cloud service providers. Furthermore, the service provider’s implementation is subject to auditing requirements, which prohibit the storage of real-time keys in local systems.
Therefore, past keys are irrecoverable to the service provider, preventing it from generating texts that falsely claim earlier timestamps.

For each text generation request within time window $t$, the system first samples a random bit sequence
\begin{equation}
R \in \{0,1\}^m,
\end{equation}
and encodes it into a redundant payload
\begin{equation}
P = \mathrm{ECC}(R),
\end{equation}
where $\mathrm{ECC}(\cdot)$ denotes an Error-Correcting Code such as BCH (Bose--Chaudhuri--Hocquenghem code). The purpose of ECC is to enlarge the redundancy distance between codewords, so that the original random sequence $R$ can still be recovered even when some payload bits are decoded incorrectly.

Next, a position allocation function maps each bit in $P$ to multiple token positions in the generated text. Denote the final payload length by $|P|$ and the generated text length by $L$. Let
\begin{equation}
\mathcal{A}: \{1,\dots,L\} \rightarrow \{1,\dots,|P|\}
\end{equation}
be the allocation function, where $\mathcal{A}(i)$ indicates which payload bit is encoded at token position $i$. For example, when $|P|=63$ and $L=630$, each payload bit is allocated to 10 token positions, so that repeated embedding improves decoding reliability.

During autoregressive generation, let $i$ denote the current token index and let $\mathrm{prefix}_i$ denote the text prefix generated before sampling the $i$-th token. The watermarking framework adopts a two-stage encoding strategy.

\paragraph{Stage I: the first $\alpha |P|$ tokens.}
For each token position $i \leq \alpha |P|$, a seed is computed using the current time key $K_t$, the random sequence $R$, and the hash of the current prefix:
\begin{equation}
S_i^{1} = \mathrm{PRF}(K_t, R, h(\mathrm{prefix}_i)),
\end{equation}
where $\mathrm{PRF}(\cdot)$ is a pseudo-random function and $h(\mathrm{prefix}_i)$ is the hash of the already generated prefix.

Given the seed $S_i^{1}$, the vocabulary $V$ is pseudo-randomly permuted:
\begin{equation}
\pi_i^{1} = \mathrm{Permute}(V; S_i^{1}),
\end{equation}
and the greenlist is defined as the first half of the permuted vocabulary:
\begin{equation}
G_i^{1} = \{\pi_i^{1}(1), \pi_i^{1}(2), \dots, \pi_i^{1}(|V|/2)\}.
\end{equation}
The remaining half forms the corresponding redlist.

Let $b_i = P_{\mathcal{A}(i)}$ be the payload bit assigned to position $i$. If $b_i = 1$, the logits of tokens in $G_i^{1}$ are increased by a bias $\delta$ during sampling; otherwise, the logits of tokens in the redlist are increased by $\delta$. In this way, the sampled token distribution is steered toward the token subset corresponding to the target payload bit.

\paragraph{Stage II: the remaining $L-\alpha |P|$ tokens.}
For token positions $i > \alpha |P|$, the procedure remains the same except that the random sequence $R$ is removed from seed generation:
\begin{align}
S_i^{2} &= \mathrm{PRF}(K_t, h(\mathrm{prefix}_i)),\\
\pi_i^{2} &= \mathrm{Permute}(V; S_i^{2}),\\
G_i^{2} &= \{\pi_i^{2}(1), \pi_i^{2}(2), \dots, \pi_i^{2}(|V|/2)\}.
\end{align}
The same position allocation function $\mathcal{A}(\cdot)$ and the same biased sampling rule are then applied. Therefore, the difference between the two stages lies only in whether $R$ participates in seed generation.

\subsection{Decoding Model}

When an intellectual-property dispute arises, the system considers all candidate time windows in the disputed period. Let
\begin{equation}
T = \{t_1, t_2, \dots, t_n\}
\end{equation}
denote the set of all possible disputed time windows. For each $t \in T$, the corresponding secret key $K_t$ is retrieved from the Hardware Security Module (HSM), and the decoder performs the following two-step recovery and verification procedure.

\paragraph{Step 1: Decode the payload and recover $R$.}
For a candidate key $K_t$, the decoder first reconstructs the second-stage seeds and greenlists for the last $L-\alpha |P|$ token positions:
\begin{equation}
S_i^{2} = \mathrm{PRF}(K_t, h(\mathrm{prefix}_i)), \qquad i > \alpha |P|,
\end{equation}
\begin{equation}
G_i^{2} = \mathrm{Greenlist}(V, S_i^{2}).
\end{equation}
Using the same position allocation function $\mathcal{A}(\cdot)$ as in encoding, the decoder groups token positions that correspond to the same payload bit. For each payload bit index $j \in \{1,\dots,|P|\}$, let
\begin{equation}
\mathcal{I}_j = \{ i \mid \mathcal{A}(i)=j,\; i > \alpha |P| \}
\end{equation}
be the set of token positions assigned to bit $P_j$ in Stage II. Then the decoder estimates $P_j$ by majority voting:
\begin{equation}
\hat{P}_j =
\begin{cases}
1, & \text{if more than half of the tokens in } \mathcal{I}_j \text{ fall into their corresponding greenlists},\\
0, & \text{otherwise}.
\end{cases}
\end{equation}
After obtaining the recovered payload $\hat{P}$, the decoder applies ECC decoding to recover the random sequence:
\begin{equation}
\hat{R} = \mathrm{ECC}_{\mathrm{Dec}}(\hat{P}).
\end{equation}

\paragraph{Step 2: Verify $\hat{R}$ and $K_t$.}
The recovered $\hat{R}$ is then used together with $K_t$ to reconstruct the first-stage seeds and greenlists for the first $\alpha |P|$ token positions:
\begin{equation}
S_i^{1} = \mathrm{PRF}(K_t, \hat{R}, h(\mathrm{prefix}_i)), \qquad i \leq \alpha |P|,
\end{equation}
\begin{equation}
G_i^{1} = \mathrm{Greenlist}(V, S_i^{1}).
\end{equation}
For each such position $i$, let $\hat{b}_i = \hat{P}_{\mathcal{A}(i)}$ denote the recovered payload bit assigned to this position. A token is counted as \emph{correctly watermarked} if it matches the intended biased subset, i.e.,
\begin{equation}
\mathrm{Match}(i)=
\begin{cases}
1, & \text{if } \hat{b}_i = 1 \text{ and token}_i \in G_i^{1},\\
1, & \text{if } \hat{b}_i = 0 \text{ and token}_i \notin G_i^{1},\\
0, & \text{otherwise}.
\end{cases}
\end{equation}
The verification score is defined as the fraction of correctly watermarked tokens in the first stage:
\begin{equation}
\mathrm{Score}(K_t,\hat{R}) = \frac{1}{\alpha |P|}\sum_{i=1}^{\alpha |P|}\mathrm{Match}(i).
\end{equation}
If
\begin{equation}
\mathrm{Score}(K_t,\hat{R}) \ge \phi,
\end{equation}
where $\phi$ is a predefined verification threshold, then $\hat{R}$ is considered valid under $K_t$, and the corresponding time window $t$ is identified as the generation time of the disputed text.

\subsection{Special Designs of TimeMark}

Our framework differs from prior watermarking methods in several important aspects.

\textbf{Hardware Security Module and one-way time-key evolution.}
For all enterprises using our time watermarking service, the key at the next time window is derived from the current key through a one-way function. Past keys are not retained by the service provider. Only the HSM stores the historical sequence of $K_t$ values as time evolves. HSMs are tamper-resistant hardware devices designed for secure cryptographic key generation, storage, and management. They are typically deployed in highly sensitive domains such as banking systems, cloud services, and government institutions, and are subject to strong regulatory and certification requirements. Under this design, the provider cannot obtain past watermarking keys and therefore cannot forge texts with backdated timestamps.

\textbf{Separating target information from watermark payload.}
In existing watermarking studies, the target information itself (e.g., timestamp) is encoded directly as the watermark payload. In contrast, our framework binds the target information to the secret key rather than to the payload, while the payload is instantiated as a fresh random bit sequence for each generation. This separation provides two security benefits. First, once time information is represented by the secret key, the provider cannot generate a text carrying an arbitrary historical timestamp unless it has access to the corresponding key. Second, because the payload is random and changes across generations, the generated texts do not exhibit statistical patterns associated with a specific payload, which fundamentally reduces the feasibility of statistical attacks.

\textbf{Short random payload with ECC-based reliable recovery.}
In practice, timestamp often has many bits and directly encoding it as payload would increase decoding difficulty. Instead, we encode a much shorter random sequence $R$ and then expand it using ECC. For example, when $R$ contains 12 bits and is encoded into a 63-bit BCH codeword, the decoder can theoretically correct up to 11 bit errors, yielding near-perfect recovery accuracy for $R$. Moreover, theoretical analysis shows that, with a suitable threshold $\phi$, an incorrect candidate key $K_t$ may still produce some decoded sequence, but its probability of passing the second-stage verification is asymptotically negligible. Therefore, the proposed design simultaneously improves decoding reliability and suppresses false acceptance under wrong keys.

\section{Experiments, Theoretical Analysis and Results}

\subsection{Experimental Setup}
We construct a prompt pool containing 20 prompts related to creative generation and report writing. In each trial, we randomly sample one prompt from this pool and use it to perform a text generation task. Each time we generate two texts under exactly the same hyperparameters: one text is generated with the proposed time watermark, and the other is generated naturally without watermark injection. This process is repeated for 800 trials in total.

For decoding, for each generated text, we consider the correct time-key together with its neighboring time-keys, resulting in a set of 5 candidate keys centered around the ground-truth key. For each candidate key, we first perform \textit{Step 1} to decode the recovered random sequence $\hat{R}$, and then perform \textit{Step 2} verification. Based on these results, we record the corresponding success rates and verification statistics.

The experiments are conducted using Qwen2.5-7B. The random payload $R$ contains 10 bits, and ECC encoding is implemented using BCH code to expand $R$ into a 63-bit payload. The verification threshold is set to $\phi = 0.65$. We set $\alpha = 5$, and the generated text length is constrained to be no shorter than $L = 945$ tokens.

\subsection{Theoretical Analysis}
Next, we provide a theoretical analysis focusing on the following aspects: (1) when $R$ contains 10 bits and is expanded to 63 bits using BCH coding, how many bit errors can be tolerated during decoding; (2) when the bias strength $\delta$ is set to 2.5, $\alpha = 5$ and $L = 945$, what is the probability of correctly recovering $R$ in Decoding \textit{Step 1}; and (3) in Decoding \textit{Step 2}, when the threshold $\phi$ is set to 0.65, what are the probabilities of the following two types of errors: first, the probability that the matching rate corresponding to the correct $K_t$ and $R$ is lower than 0.65; and second, the probability that the matching rate corresponding to an incorrect $K_t$ or $R$ exceeds 0.65.

\paragraph{(1) Error Tolerance in BCH Decoding}

Let $R \in \{0,1\}^{10}$ denote the original random sequence. In this setting, $R$ is directly treated as the 10-bit information vector of the BCH code, and the encoder outputs a 63-bit codeword
\begin{equation}
P = \mathrm{BCH}_{63,10}(R).
\end{equation}

The BCH$(63,10)$ code used here is a binary primitive BCH code of length $n=63$, dimension $k=10$, and minimum distance
\begin{equation}
d_{\min}=27.
\end{equation}
Therefore, this code has the standard parameter form
\begin{equation}
[n,k,d_{\min}] = [63,10,27].
\end{equation}

For a block code with minimum Hamming distance $d_{\min}$, the maximum number of correctable bit errors under bounded-distance decoding is
\begin{equation}
t = \left\lfloor \frac{d_{\min}-1}{2} \right\rfloor =13.
\end{equation}
This error-correction capability is a standard result in coding theory. The intuition is that a decoder can uniquely recover the transmitted codeword as long as the received word remains within the unique-decoding radius of that codeword, which is exactly half of the minimum distance between distinct codewords \citep{wicker1995error}.
Hence, BCH$(63,10)$ can theoretically correct up to 13 bit errors in the received 63-bit sequence.

This strong error-correction capability is important for the proposed watermarking framework. In Decoding Step 1, the recovered payload bits are estimated from token-level watermark signals and may contain a small number of errors. By introducing BCH$(63,10)$ coding, the framework can absorb these bit-level decoding errors and still recover the correct random sequence $R$, thereby significantly improving the reliability of time watermark detection.

\paragraph{(2) Recovery of $R$ in Decoding Step 1}

We next analyze the probability of correctly recovering $R$ in Decoding Step 1 when the watermark bias is set to the commonly used value $\delta = 2.5$.
As shown in the previous subsection, BCH$(63,10)$ can correct up to 13 bit errors.

For each token position, the vocabulary is partitioned into a greenlist and a redlist of equal size. Under the standard approximation used in green/red-list watermark analysis, adding a logit bias $\delta$ to the target subset changes the probability that the sampled token falls into the target subset from $1/2$ to
\begin{equation}
p_{\mathrm{tok}} = \frac{e^{\delta}}{1 + e^{\delta}}.
\label{eq:bias}
\end{equation}
When $\delta = 2.5$, we obtain
\begin{equation}
p_{\mathrm{tok}} = \frac{e^{2.5}}{1 + e^{2.5}} \approx 0.9241.
\label{eq:bias2}
\end{equation}
Equivalently, the probability that a token falls outside the target subset is
\begin{equation}
q_{\mathrm{tok}} = 1 - p_{\mathrm{tok}} \approx 0.0759.
\end{equation}

In our design, $L=945$ and $\alpha=5$ means that each payload bit is embedded into 10 token positions. During decoding, a payload bit is estimated by majority voting: if more than half of the 10 tokens fall into the target subset, the bit is decoded correctly; otherwise it is decoded incorrectly. Let
\begin{equation}
X \sim \mathrm{Binomial}(10, p_{\mathrm{tok}})
\end{equation}
denote the number of correctly matched tokens among the 10 positions corresponding to one payload bit. Then the probability of correctly decoding one bit is
\begin{equation}
p_{\mathrm{bit}}
= \Pr(X \ge 6)
= \sum_{k=6}^{10} \binom{10}{k} p_{\mathrm{tok}}^k (1-p_{\mathrm{tok}})^{10-k}.
\end{equation}
Substituting $p_{\mathrm{tok}} \approx 0.9241$ gives
\begin{equation}
p_{\mathrm{bit}} \approx 0.9995427.
\end{equation}
Therefore, the bit error probability is
\begin{equation}
q_{\mathrm{bit}} = 1 - p_{\mathrm{bit}} \approx 0.0004573.
\end{equation}

The 10-bit random sequence $R$ is encoded into a 63-bit BCH$(63,10)$ codeword. Since BCH$(63,10)$ can correct up to 13 bit errors, Decoding Step 1 succeeds as long as the number of incorrectly decoded payload bits does not exceed 13. Let
\begin{equation}
Y \sim \mathrm{Binomial}(63, q_{\mathrm{bit}})
\end{equation}
denote the number of erroneous decoded bits among the 63 payload bits. Then the probability of correctly recovering the original random sequence $R$ is
\begin{equation}
p_R
= \Pr(Y \le 13)
= \sum_{e=0}^{13} \binom{63}{e} q_{\mathrm{bit}}^e (1-q_{\mathrm{bit}})^{63-e}.
\end{equation}
Substituting $q_{\mathrm{bit}} \approx 0.0004573$ yields
\begin{equation}
p_R \approx 1 - 6.40 \times 10^{-34}.
\end{equation}

That is, under the above assumptions, the probability of correctly recovering $R$ in Decoding Step 1 is essentially
\begin{equation}
p_R \approx 1.
\end{equation}

It should be noted that this analysis is based on the standard independence approximation, namely that token-level decoding outcomes are treated as approximately independent Bernoulli variables with success probability $p_{\mathrm{tok}}$. While in practice the output tokens of an LLM are not independent of one another. 

More importantly, the fundamental assumption underlying Equation~\ref{eq:bias} is that the total probability mass of tokens in the greenlist is \(50\%\). This is a simplifying assumption because the output distribution of an LLM is typically concentrated on only a few tokens, which implies that the probability mass assigned to the greenlist can deviate substantially from \(50\%\). Based on data reported in prior work \cite{kirchenbauer2023watermark}, we approximately infer that in their experiments, the average total probability mass of greenlist tokens is around \(0.35\). Under this value, Equation~\ref{eq:bias2} yields a result of approximately \(87\%\). If we further calculate based on this \(87\%\), then we can get
\begin{equation}
p_R \approx 1 - 1.76 \times 10^{-19},
\end{equation}
which is still a number very close to 1.

However, the above calculation is still an estimate based on data from prior experiments, the recovery accuracy of $R$ in Step 1 still needs to be further validated through real-data experiments.

\paragraph{Matching Rate in Decoding Step 2}

We next analyze the error probabilities in Decoding Step 2 when the verification threshold is set to
\begin{equation}
\phi = 0.65.
\end{equation}
In the current setting, we use $\alpha = 5$, and the verification stage is performed over the first
\begin{equation}
\alpha |P| = 5 \times 63 = 315
\end{equation}
tokens. For each of these 315 token positions, we check whether the generated token falls into the intended biased subset determined by the corresponding key and recovered random sequence. The matching rate is therefore defined as
\begin{equation}
\mathrm{MR} = \frac{1}{315}\sum_{i=1}^{315} \mathrm{Match}(i),
\end{equation}
where $\mathrm{Match}(i) \in \{0,1\}$ indicates whether the watermark embedding at position $i$ is consistent with the expected green/red partition.

We consider two types of errors:

\begin{enumerate}
    \item the probability that the matching rate corresponding to the correct $K$ and $R$ is still smaller than $\phi=0.65$;
    \item the probability that the matching rate corresponding to an incorrect $K$ or $R$ still exceeds $\phi=0.65$.
\end{enumerate}

\textbf{Case 1: false rejection under the correct $K$ and $R$.}
When the key $K$ and the recovered random sequence $R$ are both correct, each token in Step 2 is matched against the same target subset that was favored during watermark embedding. Under the standard approximation used in watermark analysis, the probability that a token falls into the intended subset is
\begin{equation}
p_{\mathrm{match}} = \frac{e^{\delta}}{1+e^{\delta}}.
\end{equation}
With the commonly used setting $\delta = 2.5$, we have
\begin{equation}
p_{\mathrm{match}} = \frac{e^{2.5}}{1+e^{2.5}} \approx 0.9241.
\end{equation}

Let
\begin{equation}
X \sim \mathrm{Binomial}(315,\,0.9241)
\end{equation}
denote the number of matched positions among the 315 verification tokens. Then the probability of false rejection is
\begin{equation}
\Pr(\mathrm{MR}<0.65)
=
\Pr\!\left(X < 0.65 \times 315\right).
\end{equation}
Since
\begin{equation}
0.65 \times 315 = 204.75,
\end{equation}
this is equivalent to
\begin{equation}
\Pr(\mathrm{MR}<0.65)
=
\Pr(X \le 204)
=
\sum_{k=0}^{204}
\binom{315}{k}
(0.9241)^k
(1-0.9241)^{315-k}.
\end{equation}
Numerically, this probability is
\begin{equation}
\Pr(\mathrm{MR}<0.65)
\approx 1.61 \times 10^{-44}.
\end{equation}

This value is essentially zero, which means that when the correct $K$ and $R$ are used, the probability that the verification score falls below the threshold $0.65$ is negligibly small.

\textbf{Case 2: false acceptance under an incorrect $K$ or $R$.}
If either the candidate key $K$ is incorrect or the recovered random sequence $R$ is incorrect, then the reconstructed green/red partition at each token position is unrelated to the one used during encoding. In this case, for any token, the probability of accidentally matching the expected subset is approximately
\begin{equation}
p_{\mathrm{rand}} = 0.5,
\end{equation}
because the vocabulary is randomly divided into two equal halves. Let
\begin{equation}
Y \sim \mathrm{Binomial}(315,\,0.5)
\end{equation}
denote the number of accidental matches among the 315 verification tokens. Then the probability of false acceptance is
\begin{equation}
\Pr(\mathrm{MR}\ge 0.65)
=
\Pr\!\left(Y \ge 0.65 \times 315\right)
=
\Pr(Y \ge 205).
\end{equation}
Therefore,
\begin{equation}
\Pr(\mathrm{MR}\ge 0.65)
=
\sum_{k=205}^{315}
\binom{315}{k}
(0.5)^{315}.
\end{equation}
Numerically, this probability is
\begin{equation}
\Pr(\mathrm{MR}\ge 0.65)
\approx 4.76 \times 10^{-8}.
\end{equation}

The above results show that the verification threshold $\phi=0.65$ provides an extremely strong separation between correct and incorrect candidates. For the correct $K$ and $R$, the probability of obtaining a matching rate below the threshold is approximately $1.61 \times 10^{-44}$, which is effectively zero. In contrast, for an incorrect $K$ or $R$, the probability of still exceeding the threshold is only approximately $4.76 \times 10^{-8}$. Therefore, under the standard independence approximation, Decoding Step 2 can simultaneously achieve an almost-zero false rejection probability and an extremely small false acceptance probability.

\subsection{Results}

Table~\ref{tab:main_results} reports the main experimental results. Experimental results demonstrate that our method achieves perfect recovery of the generation time, while its design theoretically precludes both provider- and user-side attacks, enabling its use as reliable judicial evidence.

\begin{table}[t]
\centering
\caption{Main experimental results of the proposed time watermarking framework.}
\label{tab:main_results}
\begin{tabular}{lc}
\hline
\textbf{Metric} & \textbf{Result} \\
\hline
Number of watermarked texts with correctly identified generation time & 800 ($100\%$) \\
Number of non-watermarked texts identified as having a generation time & 0 ($0\%$) \\
Average verification score in Step 2 for watermarked texts & 0.7699 \\
Average verification score in Step 2 for non-watermarked texts & 0.4993 \\
\hline
\end{tabular}
\end{table}

\section{Conclusions and Future Directions}

This paper proposes a trustworthy time watermarking framework, which can achieve accurate watermark detection and resist both provider-side forgery and user-side forgery. The current framework still has two limitations. First, to guarantee perfect detection accuracy, it currently only supports long-text generation. Second, it assumes that users preserve the original AIGC output without manual modification. Therefore, future work can focus on reducing the required text length while maintaining trustworthiness, and improving robustness against text editing.


\bibliographystyle{informs2014}
\bibliography{main}

\begin{thebibliography}{49}
\providecommand{\natexlab}[1]{#1}
\providecommand{\url}[1]{\texttt{#1}}
\providecommand{\urlprefix}{URL }

\bibitem[{Adams et~al.(2001)Adams, Cain, Pinkas, \protect\BIBand{} Zuccherato}]{rfc3161}
Adams C, Cain P, Pinkas D, Zuccherato R (2001) Internet x.509 public key infrastructure time-stamp protocol (tsp). RFC 3161, \urlprefix\url{https://www.ietf.org/rfc/rfc3161.txt}.

\bibitem[{Bick et~al.(2026)Bick, Blandin, \protect\BIBand{} Deming}]{bick2026rapid}
Bick A, Blandin A, Deming DJ (2026) The rapid adoption of generative ai. \emph{Management Science} .

\bibitem[{Bulgakov et~al.(2024)Bulgakov, Aleshina, Smirnov, Demidov, Milyutin, \protect\BIBand{} Xin}]{bulgakov2024scalability}
Bulgakov AL, Aleshina AV, Smirnov SD, Demidov AD, Milyutin MA, Xin Y (2024) Scalability and security in blockchain networks: Evaluation of sharding algorithms and prospects for decentralized data storage. \emph{Mathematics} 12(23):3860.

\bibitem[{Chen et~al.(2024)Chen, Ma, Xu, Ma, Hu, Liu, Wu, Wang, \protect\BIBand{} Li}]{chen2024comprehensive}
Chen B, Ma L, Xu H, Ma J, Hu D, Liu X, Wu J, Wang J, Li K (2024) A comprehensive survey of blockchain scalability: Shaping inner-chain and inter-chain perspectives. \emph{arXiv preprint arXiv:2409.02968} .

\bibitem[{Christ et~al.(2024)Christ, Gunn, \protect\BIBand{} Zamir}]{christ2024undetectable}
Christ M, Gunn S, Zamir O (2024) Undetectable watermarks for language models. \emph{The Thirty Seventh Annual Conference on Learning Theory}, 1125--1139 (PMLR).

\bibitem[{{Coalition for Content Provenance and Authenticity}(2025)}]{c2pa2025}
{Coalition for Content Provenance and Authenticity} (2025) C2pa technical specification, version 2.4. \url{https://c2pa.org/specifications/specifications/2.4/specs/C2PA_Specification.html}, official technical specification for Content Credentials.

\bibitem[{{Council of Europe}(2019)}]{coe_e_evidence}
{Council of Europe} (2019) Guidelines on electronic evidence in civil and administrative proceedings. \url{https://rm.coe.int/guidelines-on-electronic-evidence-and-explanatory-memorandum/1680968ab5}, adopted by the Committee of Ministers on January 30, 2019.

\bibitem[{{Cyberspace Administration of China and Other Agencies}(2023)}]{china_genai_measures}
{Cyberspace Administration of China and Other Agencies} (2023) Interim measures for the management of generative artificial intelligence services. \url{https://www.cac.gov.cn/2023-07/13/c_1690898327029107.htm}, promulgated July 13, 2023.

\bibitem[{{Cyberspace Administration of China and Other Agencies}(2025)}]{china_ai_label_measures}
{Cyberspace Administration of China and Other Agencies} (2025) Measures for the labeling of ai-generated synthetic content. \url{https://www.cac.gov.cn/2025-03/14/c_1743654684782215.htm}, promulgated March 14, 2025.

\bibitem[{Dathathri et~al.(2024)Dathathri, See, Ghaisas, Huang, McAdam, Welbl, Bachani, Kaskasoli, Stanforth, Matejovicova et~al.}]{dathathri2024scalable}
Dathathri S, See A, Ghaisas S, Huang PS, McAdam R, Welbl J, Bachani V, Kaskasoli A, Stanforth R, Matejovicova T, et~al. (2024) Scalable watermarking for identifying large language model outputs. \emph{Nature} 634(8035):818--823.

\bibitem[{Diaa et~al.(2024)Diaa, Aremu, \protect\BIBand{} Lukas}]{diaa2024optimizing}
Diaa A, Aremu T, Lukas N (2024) Optimizing adaptive attacks against watermarks for language models. \emph{arXiv preprint arXiv:2410.02440} .

\bibitem[{{European Commission}(2025{\natexlab{a}})}]{eu_gpai_code}
{European Commission} (2025{\natexlab{a}}) The general-purpose ai code of practice. \url{https://digital-strategy.ec.europa.eu/en/policies/contents-code-gpai}, published July 10, 2025.

\bibitem[{{European Commission}(2025{\natexlab{b}})}]{eu_gpai_qa}
{European Commission} (2025{\natexlab{b}}) General-purpose ai models in the ai act -- questions \& answers. \url{https://digital-strategy.ec.europa.eu/en/faqs/general-purpose-ai-models-ai-act-questions-answers}, updated July 10, 2025.

\bibitem[{{European Union}(2014)}]{eidas_regulation}
{European Union} (2014) Regulation (eu) no 910/2014 on electronic identification and trust services for electronic transactions in the internal market (eidas regulation). \url{https://eur-lex.europa.eu/eli/reg/2014/910/oj/eng}, consolidated version available through EUR-Lex.

\bibitem[{{European Union}(2019)}]{eu_dsm_directive}
{European Union} (2019) Directive (eu) 2019/790 of the european parliament and of the council of 17 april 2019 on copyright and related rights in the digital single market. \url{https://eur-lex.europa.eu/eli/dir/2019/790/oj/eng}, official Journal of the European Union.

\bibitem[{{European Union}(2024)}]{eu_ai_act}
{European Union} (2024) Regulation (eu) 2024/1689 of the european parliament and of the council of 13 june 2024 laying down harmonised rules on artificial intelligence (artificial intelligence act). \url{https://eur-lex.europa.eu/eli/reg/2024/1689/oj/eng}, official Journal of the European Union.

\bibitem[{Fairoze et~al.(2025)Fairoze, Garg, Jha, Mahloujifar, Mahmoody, \protect\BIBand{} Wang}]{fairoze2025publicly}
Fairoze J, Garg S, Jha S, Mahloujifar S, Mahmoody M, Wang M (2025) Publicly-detectable watermarking for language models. \emph{IACR Communications in Cryptology} 1(4).

\bibitem[{Gipp et~al.(2017)Gipp, Meuschke, Beel, \protect\BIBand{} Breitinger}]{gipp2017blockchain}
Gipp B, Meuschke N, Beel J, Breitinger C (2017) Using the blockchain of cryptocurrencies for timestamping digital cultural heritage. \emph{Bulletin of IEEE Technical Committee on Digital Libraries (TCDL)} .

\bibitem[{Haber \protect\BIBand{} Stornetta(1991)}]{haber1991timestamp}
Haber S, Stornetta WS (1991) How to time-stamp a digital document. \emph{Journal of Cryptology} 3(2):99--111, \urlprefix\url{https://link.springer.com/article/10.1007/BF00196791}.

\bibitem[{Hu et~al.(2024)Hu, Chen, Wu, Wu, Zhang, \protect\BIBand{} Huang}]{hu2024unbiased}
Hu Z, Chen L, Wu X, Wu Y, Zhang H, Huang H (2024) Unbiased watermark for large language models. \emph{The Twelfth International Conference on Learning Representations}.

\bibitem[{Jiang et~al.(2025)Jiang, Wu, Boroujeny, Mark, \protect\BIBand{} Zeng}]{jiang2025stealthink}
Jiang Y, Wu C, Boroujeny MK, Mark B, Zeng K (2025) Stealthink: A multi-bit and stealthy watermark for large language models. \emph{arXiv preprint arXiv:2506.05502} .

\bibitem[{Jovanovi{\'c} et~al.(2024)Jovanovi{\'c}, Staab, \protect\BIBand{} Vechev}]{jovanovic2024watermark}
Jovanovi{\'c} N, Staab R, Vechev M (2024) Watermark stealing in large language models. \emph{arXiv preprint arXiv:2402.19361} .

\bibitem[{Kirchenbauer et~al.(2023)Kirchenbauer, Geiping, Wen, Katz, Miers, \protect\BIBand{} Goldstein}]{kirchenbauer2023watermark}
Kirchenbauer J, Geiping J, Wen Y, Katz J, Miers I, Goldstein T (2023) A watermark for large language models. \emph{International conference on machine learning}, 17061--17084 (PMLR).

\bibitem[{{Legal Information Institute}(2026{\natexlab{a}})}]{fre803}
{Legal Information Institute} (2026{\natexlab{a}}) Federal rule of evidence 803: Exceptions to the rule against hearsay. \url{https://www.law.cornell.edu/rules/fre/rule_803}, accessed April 2026.

\bibitem[{{Legal Information Institute}(2026{\natexlab{b}})}]{fre901}
{Legal Information Institute} (2026{\natexlab{b}}) Federal rule of evidence 901: Authenticating or identifying evidence. \url{https://www.law.cornell.edu/rules/fre/rule_901}, accessed April 2026.

\bibitem[{{Legal Information Institute}(2026{\natexlab{c}})}]{fre902}
{Legal Information Institute} (2026{\natexlab{c}}) Federal rule of evidence 902: Evidence that is self-authenticating. \url{https://www.law.cornell.edu/rules/fre/rule_902}, accessed April 2026.

\bibitem[{Liu et~al.(2023)Liu, Pan, Hu, Li, Wen, King, \protect\BIBand{} Yu}]{liu2023upv}
Liu A, Pan L, Hu X, Li S, Wen L, King I, Yu PS (2023) An unforgeable publicly verifiable watermark for large language models. \emph{arXiv preprint arXiv:2307.16230} .

\bibitem[{Ma \protect\BIBand{} Tsudik(2009)}]{ma2009securelogging}
Ma D, Tsudik G (2009) A new approach to secure logging. \emph{ACM Transactions on Storage} 5(1):1--21, \urlprefix\url{https://dl.acm.org/doi/10.1145/1502777.1502779}.

\bibitem[{Meng \protect\BIBand{} Chen(2022)}]{meng2022blockchain}
Meng L, Chen L (2022) A blockchain-based long-term time-stamping scheme. \emph{European Symposium on Research in Computer Security}, 3--24 (Springer).

\bibitem[{Rastogi \protect\BIBand{} Pruthi(2024)}]{rastogi2024revisiting}
Rastogi S, Pruthi D (2024) Revisiting the robustness of watermarking to paraphrasing attacks. \emph{Proceedings of the 2024 Conference on Empirical Methods in Natural Language Processing}, 18100--18110.

\bibitem[{Ren et~al.(2024{\natexlab{a}})Ren, Xu, Liu, Cui, Wang, Yin, \protect\BIBand{} Tang}]{ren2024semamark}
Ren J, Xu H, Liu Y, Cui Y, Wang S, Yin D, Tang J (2024{\natexlab{a}}) A robust semantics-based watermark for large language model against paraphrasing. \emph{Findings of the Association for Computational Linguistics: NAACL 2024}, 613--625.

\bibitem[{Ren et~al.(2024{\natexlab{b}})Ren, Xu, Liu, Cui, Wang, Yin, \protect\BIBand{} Tang}]{ren2024robust}
Ren J, Xu H, Liu Y, Cui Y, Wang S, Yin D, Tang J (2024{\natexlab{b}}) A robust semantics-based watermark for large language model against paraphrasing. \emph{Findings of the Association for Computational Linguistics: NAACL 2024}, 613--625.

\bibitem[{Siddharth(2025)}]{siddharth2025time}
Siddharth E (2025) Time-stamping in blockchain for legal evidence submission. \emph{Sci. J. Artif. Intell. Blockchain Technol} 2:34--42.

\bibitem[{{UK Government}(2024)}]{uk_copyright_ai_consultation}
{UK Government} (2024) Copyright and artificial intelligence. \url{https://www.gov.uk/government/consultations/copyright-and-artificial-intelligence/copyright-and-artificial-intelligence}, consultation published December 17, 2024.

\bibitem[{{UK Government}(2026)}]{uk_copyright_ai_progress}
{UK Government} (2026) Report on copyright and artificial intelligence. \url{https://www.gov.uk/government/publications/report-and-impact-assessment-on-copyright-and-artificial-intelligence/report-on-copyright-and-artificial-intelligence}, published March 18, 2026.

\bibitem[{{United Kingdom}(1988)}]{uk_cdpa_s3}
{United Kingdom} (1988) Copyright, designs and patents act 1988, section 3. \url{https://www.legislation.gov.uk/ukpga/1988/48/section/3}, accessed April 2026.

\bibitem[{{United States}(1976{\natexlab{a}})}]{usc17_101}
{United States} (1976{\natexlab{a}}) 17 u.s.c. § 101. \url{https://www.law.cornell.edu/uscode/text/17/101}, definitions; accessed April 2026.

\bibitem[{{United States}(1976{\natexlab{b}})}]{usc17_102}
{United States} (1976{\natexlab{b}}) 17 u.s.c. § 102. \url{https://www.law.cornell.edu/uscode/text/17/102}, subject matter of copyright: In general; accessed April 2026.

\bibitem[{{United States Courts}(2024)}]{fre2024}
{United States Courts} (2024) Federal rules of evidence. \urlprefix\url{https://www.uscourts.gov/sites/default/files/2025-02/federal-rules-of-evidence-dec-1-2024_0.pdf}, effective Dec. 1, 2024.

\bibitem[{{U.S. Copyright Office}(2025{\natexlab{a}})}]{usco_ai_overview}
{US Copyright Office} (2025{\natexlab{a}}) Copyright and artificial intelligence. \url{https://www.copyright.gov/ai/}, overview page for the Copyright and Artificial Intelligence report series.

\bibitem[{{U.S. Copyright Office}(2025{\natexlab{b}})}]{usco_ai_part2}
{US Copyright Office} (2025{\natexlab{b}}) Copyright and artificial intelligence, part 2: Copyrightability. \url{https://www.copyright.gov/ai/Copyright-and-Artificial-Intelligence-Part-2-Copyrightability-Report.pdf}, january 2025.

\bibitem[{{U.S. Copyright Office}(2025{\natexlab{c}})}]{usco_ai_part3}
{US Copyright Office} (2025{\natexlab{c}}) Copyright and artificial intelligence, part 3: Generative ai training (pre-publication version). \url{https://www.copyright.gov/ai/Copyright-and-Artificial-Intelligence-Part-3-Generative-AI-Training-Report-Pre-Publication-Version.pdf}, may 2025.

\bibitem[{Wang et~al.(2024)Wang, Yang, Chen, Zhou, Lin, Meng, Zhou, \protect\BIBand{} Sun}]{wang2024towards}
Wang L, Yang W, Chen D, Zhou H, Lin Y, Meng F, Zhou J, Sun X (2024) Towards codable watermarking for injecting multi-bits information to {LLM}s. \emph{The Twelfth International Conference on Learning Representations}, \urlprefix\url{https://openreview.net/forum?id=JYu5Flqm9D}.

\bibitem[{Wicker(1995)}]{wicker1995error}
Wicker SB (1995) \emph{Error control systems for digital communication and storage}, volume~1 (Prentice hall Englewood Cliffs).

\bibitem[{{WIPO Lex}(2020)}]{china_copyright_law}
{WIPO Lex} (2020) Copyright law of the people's republic of china. \url{https://www.wipo.int/wipolex/en/legislation/details/21065}, english version of the amended Copyright Law.

\bibitem[{{World Intellectual Property Organization}(2021)}]{wipo2021timestamp}
{World Intellectual Property Organization} (2021) Digital date-and-time-stamping: The evidentiary value and practical significance of wipo proof.

\bibitem[{Wu et~al.(2024)Wu, Hu, Guo, Zhang, \protect\BIBand{} Huang}]{wu2024resilient}
Wu Y, Hu Z, Guo J, Zhang H, Huang H (2024) A resilient and accessible distribution-preserving watermark for large language models. \emph{International Conference on Machine Learning}, 53443--53470 (PMLR).

\bibitem[{Yoo et~al.(2024)Yoo, Ahn, \protect\BIBand{} Kwak}]{yoo2024advancing}
Yoo K, Ahn W, Kwak N (2024) Advancing beyond identification: Multi-bit watermark for large language models. \emph{Proceedings of the 2024 Conference of the North American Chapter of the Association for Computational Linguistics: Human Language Technologies (Volume 1: Long Papers)}, 4031--4055.

\bibitem[{Zhang et~al.(2024)Zhang, Hussain, Neekhara, \protect\BIBand{} Koushanfar}]{zhang2024remark}
Zhang R, Hussain SS, Neekhara P, Koushanfar F (2024) $\{$REMARK-LLM$\}$: A robust and efficient watermarking framework for generative large language models. \emph{33rd USENIX Security Symposium (USENIX Security 24)}, 1813--1830.

\end{thebibliography}


\end{document}